\documentclass{elsarticle}

\usepackage[utf8]{inputenc}
\usepackage{aas_macros} 
\usepackage{graphics} 
\usepackage{amssymb}
\usepackage{amsmath}
\usepackage{xcolor}

\title{Thomas K. Gaisser, a Pioneer of Particle Astrophysics}
\author[1]{Francis Halzen} 
\author[2]{Paolo Lipari}
\address[1]{Department of Physics and \\ Wisconsin IceCube Particle Astrophysics Center \\
UW–Madison, Madison, WI 53706 USA}
\address[2]{INFN, Sezione Roma “Sapienza” \\ Piazzale Aldo Moro 2, 00185 Roma, Italy}
\date{November 2023}

\begin{document}
\maketitle
Abstract:
We describe the pioneering contributions of Thomas K. Gaisser to
the birth and development of particle astrophysics,
a new field of research at the intersection
of cosmic ray physics, astronomy, astrophysics, and particle physics
that has emerged in the last few decades.
We will especially focus on his studies of natural beams of neutrinos:
those generated by the interactions of cosmic rays in the Earth's atmosphere
and those emitted by astrophysical sources.
Tom actively participated in the discovery of these cosmic neutrinos as well. His contributions also extend to gamma-ray astronomy,
the study of the cosmic ray spectra and composition,
and the modeling of cosmic ray interactions in the atmosphere
and in astrophysical environments.
Tom invariably focused his research on the theoretical
and phenomenological problems of greatest interest at the time,
producing frameworks that transparently interpreted often complex data.
These studies have been very influential and have shaped the development
of the field.
\\

Starting in the late 1960s, the scientific life of Thomas K Gaisser spans more than five decades. This period saw the emergence and development of particle astrophysics, a new field of research at the intersection of cosmic ray physics, astronomy, astrophysics, and particle physics. If anyone could lay claim to the title of father of this new field of particle astrophysics, Tom could; he was also a true pioneer in gamma-ray and neutrino astronomy. His early career prepared him well, with research in particle and cosmic ray physics. He was a master of extracting science from the indirect information collected by air shower arrays, a skill he successfully applied in his later career to other particle astrophysics endeavors, most prominently the modeling of the atmospheric muon and neutrino fluxes. Tom's contributions to cosmic ray physics are reviewed in Ref.~\cite{Soldin:2023lbr}.

Early on, Tom studied the extensive air showers that are created when high-energy cosmic rays reach Earth. His contributions included the Gaisser-Hillas profile of longitudinal air showers developed in collaboration with Michael Hillas and the SIBYLL Monte Carlo for simulating air showers. He thus laid much of the groundwork for large experiments such as Auger and IceCube, and for how to use their data to probe fundamental questions in particle physics.

We herein emphasize Tom’s major contributions to neutrino physics. First and foremost, Tom’s research was vital for interpreting data from GeV-energy neutrino experiments using the atmospheric beam, such as IMB, Kamioka, and its next-generation successor, Super-Kamiokande. Tom provided calculations of atmospheric neutrino production that were important in establishing neutrino oscillations. With the discovery of nonvanishing neutrino masses that cannot be accommodated within the symmetries of the Standard Model, particle astrophysics made its first impactful contribution to particle physics in the era of accelerators and colliders. We will subsequently turn to Tom's major role in launching neutrino astronomy by detailed modeling of the atmospheric backgrounds, by constructing surface detectors, and by providing leadership in the construction and operation of the IceCube experiment. 

\section{Hadronic Interactions}
The importance of modeling hadronic interactions for
high-energy astrophysics is easy to understand: the highest energy cosmic rays are protons and nuclei that
can initiate hadronic interactions when they interact with ordinary matter and, in some astrophysical environments, with radiation fields.
These interactions generate an ensemble of secondaries,
and the problem is to describe with sufficient precision the
multiplicity, composition, and energy spectra of these secondary particles.
One motivation is to accurately describe the development
of the extensive air showers generated by high-energy cosmic rays in the
Earth's atmosphere. This is necessary to allow for the
reconstruction of the energy and mass number of the parent particles
from the observations of air shower detectors on the ground.
The description of hadronic interactions is also 
necessary to compute the inclusive spectra of gamma rays and neutrinos,
generated when beams of cosmic ray particles interact with some target material. This is required to evaluate the fluxes of atmospheric neutrinos generated by cosmic rays interacting in the Earth's atmosphere, and to calculate the fluxes of gamma rays and neutrinos produced in astrophysical sources, for instance, when freshly accelerated particles interact with matter inside or near a 
source that generates cosmic rays.

It is now universally accepted that, at a fundamental level,
hadronic interactions are described by Quantum Chromo Dynamics (QCD),
a gauge theory formulated in terms of quark and gluon fields. QCD theory is asymptotically free, and therefore it is possible
to accurately compute ``hard processes,'' such as high $p_\perp$ gluon-gluon 
scattering and heavy quark or Higgs boson production; however,
we are not able to compute from first principles the cross sections and the properties of the multiparticle final states for the ``soft processes''
that account for most of the interactions. For cosmic ray physics, QCD remains ``the dark side of the Standard Model.''

Tom Gaisser worked on the description of hadronic interactions and their relation to cosmic ray studies during his entire scientific life, making several important contributions. He also played a crucial role in making the community of cosmic ray physicists
aware of the critical importance of the question.
Examples of these studies include
the calculation of the atmospheric neutrino fluxes from
a knowledge of the primary cosmic ray spectra at Earth,
the study of the propagation of cosmic rays in the Galaxy via observations
of secondaries such as positrons and antiprotons~\cite{Gaisser:1982jb} generated
by inelastic collisions in interstellar space, and the modeling
of air showers, providing the tools to determine the properties of cosmic rays
\cite{Gaisser:1981qx}. These problems span a very broad range of center of mass energies, and in many
cases uncertainties in the modeling of the hadronic interactions represent the dominant source of systematic errors on the results.

Since we are not able to calculate the properties of hadronic interactions
from first principles, they must be inferred from data obtained from experiments
performed at accelerators, using as a guide a variety of phenomenological techniques.
In many cases, this requires challenging extrapolations to higher energies from the kinematical
range covered by the accelerator experiments. Tom was a master in this art.

The flow of information may also run in the opposite
direction, that is, using cosmic ray observations to obtain information on the properties of hadronic interactions.
In fact, today the cosmic ray spectra extend to energies that are
above those accessible to the highest energy man-made accelerators.
As an illustration, the collisions of a $10^{20}$~eV proton with a nucleon
at rest corresponds to a c.m. energy of 433~TeV, more than thirty times
the maximum energy (13.6~TeV) of the LHC.
Tom explored this opportunity in several of his papers.

Examples of both approaches include two papers about the behavior of the proton-air cross section with energy.
In the 1970s, experiments at the CERN ISR collider demonstrated that the
$pp$ cross section grows with energy. This result is obviously relevant for the interpretation of air shower
experiments, and Tom promptly wrote papers~\cite{Gaisser:1974xk,Gaisser:1977pi}
that discussed the implications of the ISR results for the 
determination of the cosmic ray energy spectrum. In collaboration with Gaurang Yodh, Tom also discussed how the energy dependence of the $p$--air cross section could be
inferred from cosmic ray observations~\cite{Gaisser:1974sa}.
Note that the program to evaluate the $p$--air and the $pp$ cross sections
at super-LHC-energy is currently pursued by air shower experiments,
in particular at the highest energies by the Pierre Auger Observatory
\cite{PierreAuger:2012egl}
and the Telescope Array\cite{Abbasi:2015fdr}.

Another example of the use of air shower observations to infer the properties
of hadronic interactions is the estimate of the size of Feynman scaling violations.
In 1969, Richard Feynman~\cite{Feynman:1969ej} proposed a scaling law for the longitudinal
momentum distribution of the mesons created in high-energy inelastic
interactions, referred to as ``Feynman scaling.'' It allows for the extrapolation of accelerator data to arbitrarily high energy, and it therefore has very important implications
for cosmic ray studies. However, it was soon understood, including by Feynman himself~\cite{Feynman:1969ej}, that this scaling is violated.
A paper by Gaisser and Maurer~\cite{Gaisser:1972pc}
discussed how air shower observations indicated that the multiplicity of secondaries
in hadronic collisions was indeed growing more rapidly than the logarithmic
behavior predicted by Feynman scaling.

In 1980, Tom, together with Gaurang Yodh, 
wrote an influential review~\cite{Gaisser:1980zp}
summarizing the status of the problem of modeling hadronic
interactions for cosmic ray studies. It remains today an authoritative source discussing the systematic uncertainties in 
interpreting air shower experiments. Tom Gaisser pursued
these studies for the following four decades.
A major ``product'' of these studies has been the development
of detailed Monte Carlo codes for the simulation of hadronic collisions and the development of air showers.
The importance of these computer codes has been recognized for decades because a meaningful comparison of the results of different experiments requires the use of the same Monte Carlo codes for the simulation of air showers. This task became easier
 with the development in the 1990s of CORSIKA~\cite{Heck:1998vt} a Monte Carlo code for
 air shower simulation intended for general use that also includes the possibility to use different hadronic interaction models. 

Tom Gaisser and his collaborators played an important role
in supporting and stimulating these developments, including by constructing
and developing the SIBYLL model
\cite{Fletcher:1994bd,Ahn:2009wx,Fedynitch:2018cbl,Riehn:2019jet}. This (open source) code was made public,
included in CORSIKA air shower Monte Carlo program, and has been extensively used for the interpretation of air shower observations in a broad energy range that extends from $10^{12}$ to $10^{20}$~eV.
The Monte Carlo codes developed for cosmic rays studies
are complementary to those developed for
accelerators studies in the sense that their emphasis
is not on the small part of the cross section where hard parton-parton
scattering occurs but focused on the ``minimum bias'' interactions,
with special attention to the forward fragmentation region
that is most important for shower development.

\section{Atmospheric Neutrinos}

The study of atmospheric neutrinos has been a topic of great importance
in Tom Gaisser's research, and it is in this field that he made some of his most important scientific contributions.

Atmospheric neutrinos are created in the weak decays of
secondary particles produced by cosmic rays in the Earth's atmosphere.
The main channel of atmospheric neutrino production is the chain decay
of charged pions: $\pi^+ \to \mu^+ ~ \nu_\mu$ followed by
$\mu^+ \to e^+ ~ \nu_e ~ \overline{\nu}_\mu$, and the charge conjugate mode.
Subdominant contributions at the level of $10$--20\% are generated by the production and decay of kaons, in modes such as
$K^+ \to \pi^0 + e^+ + \nu_e$ or
$K_L \to \pi^\pm + e^\mp + \overline{\nu}_e (\nu_e)$.

The existence of atmospheric neutrinos had been anticipated during the extraordinary, one could say ``heroic,''
decade (1937--1947) of studies of cosmic ray radiation
that saw the discovery of the muon and the charged pions 
and the observation of their decay modes.
By the end of the 1940s, the study of cosmic ray interactions had established that they were a source of large fluxes of neutrinos, and the main channels of atmospheric neutrino production had
been identified. The detection of this flux appeared to be very challenging, but this did not discourage physicists
to investigate methods to detect it. Two experiments located in deep mines in India and South Africa
\cite{Achar:1965ova,Reines:1965qk}
obtained the first direct observations of the flux of atmospheric 
$\nu_\mu$ and $\overline{\nu}_\mu$ by detecting the muons generated
by the charged current interactions of atmospheric neutrinos in the rock surrounding the detector.
The long range of the secondary muons allows the use of a large volume of rock as target for the neutrino interactions,
increasing significantly the event rate relative to events where the neutrinos interact inside
the detection volume.

The development of Grand Unified Theories in the 1970s
\cite{Pati:1973rp,Pati:1973uk,Georgi:1974sy}, along with their fascinating prediction of proton decay with a lifetime possibly observable by
ambitious but realistic projects~\cite{Georgi:1974yf}, stimulated the design and construction of very large mass underground detectors. These were also perfectly suited to study the interactions of atmospheric neutrinos in the GeV-energy range. This energy range, unfortunately in the vicinity of the proton mass,
is in fact where, folding a rapidly falling neutrino spectrum with a cross section growing with energy, atmospheric neutrino interactions are most frequent and become a background in the search for proton decay. The calculation of atmospheric neutrino fluxes thus became a critical task. From the beginning, Tom Gaisser became a pioneer and a leader in these studies, constantly
refining the modeling over several decades~\cite{Gaisser:1983vc,Barr:1988rb,Barr:1989ru,Agrawal:1995gk,Lipari:1998cm,
Gaisser:2001sd,Gaisser:2002zv,Barr:2004br,Barr:2006it}.

The task of predicting atmospheric neutrino fluxes
acquired a greater importance when observations
suggested the possibility that these fluxes may be modified by
flavor oscillations. Confronting precise predictions that did not include oscillations with the data became essential to establish the existence of this new phenomenon.

The first method for estimating the atmospheric
neutrino flux was to derive the flux directly from the measured muon flux.
The $\mu$ and $\nu$ fluxes are related because they are generated in the decay of the same parent particles. However, this approach has its limits, as muons are unstable, with a lifetime of 2.2~$\mu$s corresponding to a decay length $\ell_\mu \simeq 6.2~E_{\rm GeV}$~km. Because the atmosphere has a thickness on the order of 20~km,
this implies that only muons with energies exceeding a few GeV reach the ground before decay, while lower-energy muons will decay before reaching the ground.
It is relatively straightforward to extrapolate from the muon spectrum to the spectrum of $(\nu_\mu + \overline{\nu}_\mu)$ for energies $E \gtrsim 10$~GeV, because each $\mu^\pm$ is accompanied
by a $\nu_\mu(\overline{\nu}_\mu)$ created in the same $\pi^\pm$ decay. 
On the other hand, the method has limited validity
for the lower energies that are relevant for estimating
the $p$-decay background. The ``corresponding'' muons decay in air.
The extrapolation method is also not viable for estimating 
the $\nu_e$ and $\overline{\nu}_e$ fluxes.

The method followed by Tom Gaisser to estimate the
atmospheric neutrino fluxes was to perform a ``direct calculation,''
starting from measurements of the primary cosmic ray flux and
calculating the neutrino flux produced during shower development. Such a direct calculation of the atmospheric neutrino fluxes as a function of flavor, energy, and direction is based on the following elements: 
(a) a description of the primary cosmic ray fluxes,
(b) a model for the hadronic interactions of
relativistic protons and nuclei with the oxygen and nitrogen nuclei in the air,
(c) the description of the weak decays of the unstable parent particles produced in these interactions, $\mu^\pm$, $\pi^\pm$, $K^\pm$, $\ldots$,
and (d) a calculation scheme to put together the elements
listed above.
In addition, the calculation of the observable neutrino
event rates requires a precise description of the neutrino cross section. Tom Gaisser developed all of the above elements of the atmospheric
neutrino calculations.
The description of the initial cosmic ray spectra, their energy dependence, and mass composition is obviously a fundamental problem
for cosmic ray astrophysics, and Tom had studied this problem extensively. Also, as previously discussed, the modeling of hadronic interactions is a topic of great importance not only for the physics of atmospheric neutrinos, but also in a wide range of problems in high-energy astrophysics, and was always central in Tom's research.

Additionally, Tom and his collaborators were the first
\cite{Barr:1988rb} to include muon polarization
in the decay chains that generate the neutrinos. Polarization contributes non-negligible corrections that had been overlooked in the first calculations of the atmospheric neutrino fluxes.

Where observations of the neutrino flux are concerned, the description of the neutrino-nucleon cross sections, especially at the lower energies, is a nontrivial problem to which Tom contributed~\cite{Gaisser:1986bv}. It is still a source of systematic uncertainties for long baseline neutrino experiments today.

Different computational methods were used to
obtain the atmospheric neutrino fluxes. Tom preferred to use semi-analytic methods~\cite{Gaisser:2001sd} that
allow for a good understanding of the results and their systematic uncertainties, but the nature of the problem, eventually, required performing very detailed Monte Carlo calculations. The calculations of the atmospheric neutrino flux performed before 1999
made the simplifying assumption that the neutrino is collinear with the parent particle. In a Monte Carlo calculation, this can be implemented by rotating the 3-momenta of particles in the final state of an interaction (or decay) to align them to the momentum of
the projectile (parent) particle.
The use of this approximation allows saving a very large factor
in computation time. A fully 3D Monte Carlo calculation is very inefficient, because only a small fraction of the neutrinos arrives close to the detector.

In the end, the availability of modern computers made a 3D calculation possible.
The first study \cite{Battistoni:1999at} was performed using the Fluka Monte Carlo code,
showing that the 3D effects can be significant when the average angle between the
neutrinos and the primary particle is large (as also discussed in Ref.~\cite{Lipari:2000wu}).
Soon afterward, Tom and his collaborators \cite{Barr:2004br} obtained detailed predictions, which also included a 3D treatment that is significant at low energy ($E \lesssim 1$~GeV).

The large mass detectors designed
to search for proton decay, soon observed 
neutrino interactions inside the detector's fiducial volume.
In 1986, two water Cherenkov detectors, IMB~\cite{Haines:1986yf}
and Kamiokande~\cite{Nakahata:1986zp}, observed the first hints of an ``anomaly'' in the $\mu/e$ flavor ratio for contained events. The ``anomaly'' turned into a ``hint for oscillations''
when the Kamiokande collaboration released
additional data~\cite{Hirata:1992ku,Fukuda:1994mc}
with larger exposures, which strengthened the
case for the flavor oscillations.
The ``hint'' became ``evidence'' for oscillations when 
the new Super-Kamiokande experiment released its first 1.5 years
of data with a 33-kton-yrs exposure at the Neutrino'98 conference in Toyama~\cite{Fukuda:1998mi,Kajita:2000mr}.
The improved significance resulted not only from smaller statistical errors but also from the broader range in energy that the larger exposure made possible. This exposed the characteristic dependence of the neutrino pathlength and energy expected for flavor oscillations. Additional support for the flavor oscillation hypothesis emerged from the study of the zenith angle dependence of upgoing muons from the decay of neutrinos of muon flavor, a result independently confirmed by MACRO~\cite{Ambrosio:1998wu}.

The 2015 Nobel Prize in Physics was awarded to the spokesperson of the Super-Kamiokande experiment, Takaaki Kajita, and to Arthur McDonald, the spokesperson of the SNO solar neutrino detector, ``for the discovery of neutrino oscillations, which shows that neutrinos have mass.'' The discovery over the 1986--1998 period that flavor oscillations
distort the spectra of atmospheric neutrinos is a fascinating and very instructive example of the
``process of discovery.'' While the experimentalists obtained measurements with decreasing statistical and systematic errors, the
theoretical predictions were constantly refined, with Tom playing an essential role.
The most accurate and better-controlled predictions used in the interpretation of the data of the experiments were obtained by Tom and his collaborators, the ``Bartol model,'' and by a Japanese group originally formed by Kajita together with
Honda, Kasahara, and Midorikawa~\cite{Honda:1995hz,Honda:2006qj}.
An authoritative review of the subject was written by Tom, together with Morihiro Honda~\cite{Gaisser:2002jj}.

The calculation of the atmospheric neutrino flux at very high energy
(from $E \gtrsim 10$~TeV up to several PeV)
is important also because in this energy range the atmospheric flux
is the foreground to the emerging 
signal of neutrinos emitted by astrophysical sources (that have a harder energy spectrum).

An important source of uncertainty in this calculation is the modeling of the spectrum
and composition of the primary cosmic rays that generate the atmospheric neutrinos.
In the relevant energy range, the primary CR spectrum and composition can
only be measured by extensive air shower experiments, and this results in large
uncertainties (and also some discrepancies between different measurements)
because of the systematic errors associated with modeling of hadronic cascades.
The situation is particularly difficult (and interesting) because it is in this range that the
all--particle CR spectrum exhibits the softening feature known as the ``knee,''
where the composition is also rapidly changing, and because 
somewhere above the ``knee'' one also expects the transition from
a CR flux generated by Galactic sources
to a flux dominated by the extragalactic contribution.

Modeling the CR spectra and composition at very high energy
was therefore a very important subject of study for
Tom Gaisser, in part because of its relevance in the construction of precise predictions
for the atmospheric neutrino fluxes and because of its fundamental significance for cosmic ray astrophysics.

The studies by Tom on this problem resulted in publication of several 
parametrizations of the CR spectra at very high energy (from $10^{14}$ to $10^{20}$~eV)
obtained from an analysis of all existing measurements
(combined, taking into account their systematic uncertainties),
and using functional forms motivated by physical ideas about the properties of CR sources.
A well-known result that emerged from these studies is the set of parametrizations of the CR spectra 
 \cite{Gaisser:2013bla} (with Todor Stanev and Serap Tilav as coauthors)
that has been used by a large number (several hundred) of subsequent publications.

\section{High-Energy Neutrino Astronomy}

The understanding that cosmic rays may be generated in discrete astrophysical sources emerged gradually, and with it the idea that sites where cosmic rays are produced should be sources of high-energy neutrinos, 
generated by the same mechanism
that creates the atmospheric $\nu$ flux. 
These ideas made a powerful impact on astrophysics in the 1980s because of puzzling observations of Cygnus X-3 and
the explosion of supernova SN1987A in the Large Magellanic Cloud.

Tom’s first papers on astrophysical neutrinos were inspired by the pioneering attempt to construct the DUMAND neutrino telescope off the coast of the Big Island of Hawaii. Using his skills as a particle physicist, he calculated the high-energy neutrino and antineutrino cross sections~\cite{Gaisser:1977yra} a decade before the “early” papers evaluating the deep inelastic cross section for the interactions of neutrinos in water, which we still reference today. His next paper testifies to his roots in particle physics, with a calculation of the production of the Higgs boson in muon number violating processes~\cite{Gaisser:1978mj}. He also suggested the construction of an air shower array near the DUMAND site~\cite{Elbert:1980sq}, an idea that would be repeatedly and very successfully implemented with the development of neutrino astronomy in Antarctica.

The early origins of what is now referred to as particle astrophysics are partially rooted in two momentous events: the 1978 Topical Conference on Cosmic Rays and Particle Physics~\cite{Gaisser:1979hp} and the “discovery” of gamma rays and muons from Cygnus X-3. One year after completing the transfer of the Bartol Institute from its location on the Swarthmore campus near Philadelphia to the University of Delaware, Tom organized a meeting uniting cosmic ray and particle physicists to discuss common interests. One need not be a historian to track early initiatives in the new discipline to this gathering of minds---among the particle physicists James Bjorken, Carlo Rubbia, David Cline, Gordon Kane, and César Lattes, and Saburo Miyake, Kiyoshi Niu, Michael Hillas, and George Cassidy among many others on the cosmic ray side.

In 1983, a paper by Samorski and Stamm~\cite{Samorski:1983zm}
reported on ``the first experimental evidence for a clearly identified gamma-ray point source''
obtained with an air shower detector located in Kiel (Germany).
The signal was observed for energies in the range of 1--20~PeV ($10^{15}$~eV) and 
from a direction consistent with the position of the binary system Cygnus X-3, and it appeared to be time-modulated with the 4.8~hours orbital period of the system.

Several decades after this publication and a sequence of other detections, none totally compelling, observations by significantly more sensitive detectors and analyses based on more sophisticated and robust statistical methods, we can conclude that these experimental results were all incorrect, or perhaps associated to a time interval where the emission from Cygnus X-3 was
much more powerful than what is observed today~\cite{Abdo:2009kfa}. However, the fleeting reality of these exciting results had a profound
impact on the field of high-energy astrophysics.
On the experimental side, it stimulated the design and construction of better air shower detectors,
such as the Chicago Air Shower Array (CASA), which together with Michigan Muon Array (MIA)
operated in the 1990s, obtaining stringent limits on the emission from Cygnus X-3
and other possible gamma ray sources.
On the theoretical side, the observations of Cygnus X-3 stimulated more realistic modeling of the anticipated fluxes of gamma rays and neutrinos from potential cosmic ray sources.

In fact, for one decade, Cygnus X-3 became very much the focus of the new discipline, drawing more particle physicists into what was referred to at the time as non-accelerator particle physics. For astronomers, Cygnus X-3 is a high-mass X-ray binary and one of the stronger binary X-ray sources in the sky. It is believed to be a compact object, a neutron star or black hole, in a binary system that is pulling in a stream of gas from an ordinary star companion. For the particle physicist, it is a particle beam powered by a compact object aimed at a star that is the target for producing gamma rays and neutrinos. With others, Tom jumped on the fact that the observations of TeV gamma rays as well as muons, presumably produced by neutrinos, could not be accommodated by Standard Model physics~\cite{Gaisser:1983cj}. Gamma rays can be separated in cosmic ray experiments by the fact that they initiate a purely electromagnetic shower except for the photoproduction of some pions in the development of the air shower; their decay is the source of a small muon component. One possibility was that the abundant production of muons required to accommodate the observations was associated with a ``new physics" TeV threshold in the photoproduction cross section, a threshold that was anticipated by the particle community and inspired the construction of the LHC. Though the evidence faded with improved measurements and with a better understanding of statistical trials in the data analysis, the idea nevertheless took hold that one could do particle physics and search for physics beyond the Standard Model with experiments of modest cost at energies often exceeding those of earthbound accelerators. In this context, Tom started to write inspiring reviews on the prospects for neutrino astronomy, which he presented at the 2nd~\cite{BaldoCeolin:1990hg} and 6th Venice Neutrino Telescopes Workshop. These were early versions of the 1995 Physics Reports~\cite{Gaisser1995a}, which became the most widely cited review in the field. At this opportune time, Tom organized the Particle Astrophysics in Antarctica Meeting at the Bartol Institute, building on the historic role it played in Antarctic science with Martin Pomerantz~\cite{Gaisser:1979hp}. The meeting transformed neutrino astronomy in Antarctica from an idea into a project.

Then supernova 1987A exploded. Supernova 1987A and its observation by neutrino detectors did create the realistic and much-explored opportunity to do particle physics with the data from an astronomical source. Tom did not pass up the opportunity~\cite{Gaisser:1987fn}, but he also drew attention to the fact that 1987A might become a source of GeV-energy gamma rays and neutrinos long after it faded~\cite{Gaisser:1987wm}. Shockwaves may accelerate protons inside a young supernova shelll, and subsequently energetic pions can be produced that decay into photons and neutrinos of much higher energy than the deleptonization and thermal neutrinos initially produced. Their observation will be possible when the next Galactic supernova explodes, relying on a wealth of 21st century instrumentation. At the time, the air shower array at the South Pole provided a unique opportunity because of its southern location~\cite{Gaisser:1987fn}.

What is missing in this narrative so far is Tom’s knack for occasionally producing or contributing to isolated papers that are real gems. The paper with Gary Steigman and Serap Tilav on searching for dark matter particles trapped in the sun~\cite{Gaisser:1986ha} comes to mind, as well as the papers on muon-poor gamma ray astronomy applied to Jim Cronin’s effort to turn the CASA-MIA air shower array in Utah into a gamma ray telescope~\cite{Gaisser:1991tz}. Also, the highly imaginative paper with Todor Stanev and David Seckel on the observation at Earth of solar atmospheric neutrinos deserves being singled out~\cite{Seckel:1990rk}. Even today these papers inspire active searches for high-energy neutrinos originating in the sun with GeV energy and above. These have resulted in the best limits on the cross section for dark matter interacting with ordinary matter via spin-dependent interactions~\cite{Arguelles:2019jfx}. The search for solar ``atmospheric" neutrinos is closing in on their original predictions, while the corresponding photons have been recently observed by the NASA Fermi satellite~\cite{Ajello:2021agj}. 

Starting with the early deployments of AMANDA, neutrino astronomy became Tom’s focus, initially shared with his pioneering calculations of atmospheric neutrino fluxes~\cite{Gaisser:1983cj,Gaisser:1983vc}. For AMANDA and IceCube observations, atmospheric neutrinos represent both signal and background depending on the science, and, maybe more important, they represent an opportunity to calibrate the detectors over a wide energy range~\cite{Gaisser:1985cm,Gaisser:1998hb}. Tom led the development of the tools to evaluate the atmospheric neutrino flux, including uncertainties~\cite{Barr:2007fza}. As a founding member of the IceCube Collaboration established in 2005, Tom was a leader whom everyone could count on. He was gracious and provided encouragement to many young scientists. He served the team in many ways, including as IceCube’s spokesperson between 2007 and 2011.

Tom was the soul of IceTop, the observatory’s surface air shower array built for calibrating IceCube but also devoted to cosmic-ray physics. Although a theorist, Tom took on the experimental task of building IceTop with gusto and participated in every season of IceCube’s construction. For several years he traveled to Antarctica, staying there for weeks at a time to participate in building the surface array. He delighted in the hard physical labor and the camaraderie of everyone, including mechanics, bulldozer drivers, and technicians engaged in the project. IceTop and IceCube mapped for the first time the cosmic-ray anisotropy in the Southern Hemisphere and performed precision measurements of the cosmic-ray energy spectrum from the ``knee" to the ``ankle" in the spectrum. As an IceCube member, Tom also became an ambassador of Antarctic science in large part through a blog documenting his team’s expeditions to the South Pole. In recognition of his work with IceCube, an area in Antarctica was named Gaisser Valley in 2005.

Given his pioneering role, from the first SPASE air shower array deployed at the South Pole to his leadership in IceCube from the time of construction to its present discoveries, it is appropriate to briefly summarize IceCube’s results. In its first decade of operation, IceCube collected on the order of one million neutrinos, mostly of atmospheric origin. Among these, it discovered neutrinos of TeV-PeV energy originating beyond our Galaxy, providing us with the only unobstructed view of the cosmic accelerators that power the highest energy radiation reaching us from the extreme universe~\cite{Aartsen:2013jdh}. Increasingly precise measurements of their spectrum using multiple methodologies revealed two surprises. First, unlike the case for all wavebands of light, the contribution to the cosmic neutrino flux from our own Galaxy is only at the 10\% level~\cite{Aartsen:2014gkd}. Second, the expected flux of gamma rays from the decay of the neutral pions accompanying the charged pions that decay into cosmic neutrinos exceeds the total extragalactic flux observed by gamma ray detectors. It implies that the targets in which the cosmic accelerators produce neutrinos are opaque to gamma rays. This has been confirmed by the identification of the first neutrino sources.

The self-veto of atmospheric muon neutrinos has been a critical tool for identifying cosmic neutrinos in the overwhelming atmospheric neutrino background. The latter are identified by accompanying muons from the neutrino decay, or muons produced in the same air shower as the neutrino. Even though the original idea was not his, it represents one more example demonstrating that Tom's research invariably focused on the theoretical and phenomenological aspects of the measurements that were of great interest at the time. Be it air showers, anti-proton production and propagation, lepton fluxes (and their dependence on the K/pi ratio, atmospheric parameters, etc.), or astrophysical neutrino and gamma-ray fluxes, he always produced estimates that were extremely helpful for interpreting data.

After 10 years of accumulated statistics, the active galaxy NGC 1068 has been associated with the hottest spot in the neutrino sky map. It is also the dominant source in a search at the positions of 110 preselected high-energy gamma-ray sources. At the location of NGC 1068, we observe an excess of 79$^{+22}_{-20}$
neutrinos with TeV energies~\cite{IceCube:2022der}. Additionally, IceCube has found evidence for the active galaxies PKS 1424+240, TXS 0506+056, and NGC 4151. TXS 0506+056 had already been identified as a neutrino source in a multimessenger campaign triggered by a neutrino of 290 TeV energy, IC170922~\cite{IceCube:2018dnn}, and by the independent observation of a neutrino burst from this source in archival IceCube data in 2014~\cite{IceCube:2018cha}. The observations point to active galaxies opaque to gamma rays, with the obscured dense cores near the supermassive black holes emerging as the sites where neutrinos originate, typically within 10-100 Schwarzschild radii. IceCube is thus closing in on the resolution of the century-old problem of where cosmic rays originate. 

The background of atmospheric neutrinos provides IceCube with a high-statistic sample to study the oscillations of neutrinos. IceCube's measurements of the so-called atmospheric neutrino parameters have reached a precision similar to what has been achieved by accelerator experiments~\cite{yu2023recent}. These measurements are performed at higher energies and will be further improved after the deployment of the IceCube Upgrade in the 2025-26 South Pole summer.

All of the important advances in particle astrophysics discussed above have followed in large part from the foundations laid by Tom Gaisser's work. And for that, 
Tom Gaisser's name will forever be associated with neutrino physics that takes advantage of the atmospheric neutrino beam and with the birth of high-energy neutrino astronomy.
 
\bibliographystyle{elsarticle-num}
\bibliography{bib}

\begin{thebibliography}{10}
\expandafter\ifx\csname url\endcsname\relax
  \def\url#1{\texttt{#1}}\fi
\expandafter\ifx\csname urlprefix\endcsname\relax\def\urlprefix{URL }\fi
\expandafter\ifx\csname href\endcsname\relax
  \def\href#1#2{#2} \def\path#1{#1}\fi

\bibitem{Soldin:2023lbr}
D.~Soldin, P.~A. Evenson, H.~Kolanoski, A.~A. Watson, {Cosmic-Ray Physics at the South Pole} (11 2023).
\newblock \href {http://arxiv.org/abs/2311.14474} {\path{arXiv:2311.14474}}.

\bibitem{Gaisser:1982jb}
T.~K. Gaisser, B.~G. Mauger, {Calculation of Cosmic Ray anti-proton ratio}, Astrophys. J. Lett. 252 (1982) L57--L59.
\newblock \href {https://doi.org/10.1086/183719} {\path{doi:10.1086/183719}}.

\bibitem{Gaisser:1981qx}
T.~K. Gaisser, T.~Stanev, P.~Freier, C.~J. Waddington, {Nucleus-nucleus Collisions and Interpretation of Cosmic Ray Cascades Above 100-tev}, Phys. Rev. D 25 (1982) 2341--2350.
\newblock \href {https://doi.org/10.1103/PhysRevD.25.2341} {\path{doi:10.1103/PhysRevD.25.2341}}.

\bibitem{Gaisser:1974xk}
T.~K. Gaisser, C.~J. Noble, G.~B. Yodh, {Contribution of an Increasing Proton Proton Cross-Section to Steepening of the Cosmic Ray Energy Spectrum}, J. Phys. G 1 (1975) L9, [Erratum: J.Phys.G 1, 789 (1975)].
\newblock \href {https://doi.org/10.1088/0305-4616/1/1/003} {\path{doi:10.1088/0305-4616/1/1/003}}.

\bibitem{Gaisser:1977pi}
T.~K. Gaisser, F.~Siohan, G.~B. Yodh, {An Estimation of the Primary Proton Spectrum Between 10**12-eV and 10**14-eV}, J. Phys. G 3 (1977) L241--L244.
\newblock \href {https://doi.org/10.1088/0305-4616/3/10/003} {\path{doi:10.1088/0305-4616/3/10/003}}.

\bibitem{Gaisser:1974sa}
T.~K. Gaisser, G.~B. Yodh, {Energy dependence of sigma(p-air) up to 20 tev}, Nucl. Phys. B 76 (1974) 182--188.
\newblock \href {https://doi.org/10.1016/0550-3213(74)90147-3} {\path{doi:10.1016/0550-3213(74)90147-3}}.

\bibitem{PierreAuger:2012egl}
P.~Abreu, et~al., {Measurement of the proton-air cross-section at $\sqrt{s}=57$ TeV with the Pierre Auger Observatory}, Phys. Rev. Lett. 109 (2012) 062002.
\newblock \href {http://arxiv.org/abs/1208.1520} {\path{arXiv:1208.1520}}, \href {https://doi.org/10.1103/PhysRevLett.109.062002} {\path{doi:10.1103/PhysRevLett.109.062002}}.

\bibitem{Abbasi:2015fdr}
R.~U. Abbasi, et~al., {Measurement of the proton-air cross section with Telescope Array\textquoteright{}s Middle Drum detector and surface array in hybrid mode}, Phys. Rev. D 92~(3) (2015) 032007.
\newblock \href {http://arxiv.org/abs/1505.01860} {\path{arXiv:1505.01860}}, \href {https://doi.org/10.1103/PhysRevD.92.032007} {\path{doi:10.1103/PhysRevD.92.032007}}.

\bibitem{Feynman:1969ej}
R.~P. Feynman, {Very high-energy collisions of hadrons}, Phys. Rev. Lett. 23 (1969) 1415--1417.
\newblock \href {https://doi.org/10.1103/PhysRevLett.23.1415} {\path{doi:10.1103/PhysRevLett.23.1415}}.

\bibitem{Gaisser:1972pc}
T.~K. Gaisser, R.~H. Maurer, {Extensive air showers and Feynman scaling above 1000 GeV}, Phys. Lett. B 42 (1972) 444--448.
\newblock \href {https://doi.org/10.1016/0370-2693(72)90103-7} {\path{doi:10.1016/0370-2693(72)90103-7}}.

\bibitem{Gaisser:1980zp}
T.~K. Gaisser, G.~B. Yodh, {PARTICLE COLLISIONS ABOVE 10-TeV AS SEEN BY COSMIC RAYS}, Ann. Rev. Nucl. Part. Sci. 30 (1980) 475--542.
\newblock \href {https://doi.org/10.1146/annurev.ns.30.120180.002355} {\path{doi:10.1146/annurev.ns.30.120180.002355}}.

\bibitem{Heck:1998vt}
D.~Heck, J.~Knapp, J.~N. Capdevielle, G.~Schatz, T.~Thouw, \href{http://inspirehep.net/record/469835/files/FZKA6019.pdf}{{CORSIKA: A Monte Carlo code to simulate extensive air showers}}, Tech. Rep. FZKA 6019 (1998).
\newline\urlprefix\url{http://inspirehep.net/record/469835/files/FZKA6019.pdf}

\bibitem{Fletcher:1994bd}
R.~S. Fletcher, T.~K. Gaisser, P.~Lipari, T.~Stanev, {SIBYLL: An Event generator for simulation of high-energy cosmic ray cascades}, Phys. Rev. D 50 (1994) 5710--5731.
\newblock \href {https://doi.org/10.1103/PhysRevD.50.5710} {\path{doi:10.1103/PhysRevD.50.5710}}.

\bibitem{Ahn:2009wx}
E.-J. Ahn, R.~Engel, T.~K. Gaisser, P.~Lipari, T.~Stanev, {Cosmic ray interaction event generator SIBYLL 2.1}, Phys. Rev. D80 (2009) 094003.
\newblock \href {http://arxiv.org/abs/0906.4113} {\path{arXiv:0906.4113}}, \href {https://doi.org/10.1103/PhysRevD.80.094003} {\path{doi:10.1103/PhysRevD.80.094003}}.

\bibitem{Fedynitch:2018cbl}
A.~Fedynitch, F.~Riehn, R.~Engel, T.~K. Gaisser, T.~Stanev, {Hadronic interaction model sibyll 2.3c and inclusive lepton fluxes}, Phys. Rev. D 100~(10) (2019) 103018.
\newblock \href {http://arxiv.org/abs/1806.04140} {\path{arXiv:1806.04140}}, \href {https://doi.org/10.1103/PhysRevD.100.103018} {\path{doi:10.1103/PhysRevD.100.103018}}.

\bibitem{Riehn:2019jet}
F.~Riehn, R.~Engel, A.~Fedynitch, T.~K. Gaisser, T.~Stanev, {Hadronic interaction model Sibyll 2.3d and extensive air showers}, Phys. Rev. D 102~(6) (2020) 063002.
\newblock \href {http://arxiv.org/abs/1912.03300} {\path{arXiv:1912.03300}}, \href {https://doi.org/10.1103/PhysRevD.102.063002} {\path{doi:10.1103/PhysRevD.102.063002}}.

\bibitem{Achar:1965ova}
C.~V. Achar, et~al., {Detection of muons produced by cosmic ray neutrinos deep underground}, Phys. Lett. 18 (1965) 196--199.
\newblock \href {https://doi.org/10.1016/0031-9163(65)90712-2} {\path{doi:10.1016/0031-9163(65)90712-2}}.

\bibitem{Reines:1965qk}
F.~Reines, M.~F. Crouch, T.~L. Jenkins, W.~R. Kropp, H.~S. Gurr, G.~R. Smith, J.~P.~F. Sellschop, B.~Meyer, {Evidence for high-energy cosmic ray neutrino interactions}, Phys. Rev. Lett. 15 (1965) 429--433.
\newblock \href {https://doi.org/10.1103/PhysRevLett.15.429} {\path{doi:10.1103/PhysRevLett.15.429}}.

\bibitem{Pati:1973rp}
J.~C. Pati, A.~Salam, {Is Baryon Number Conserved?}, Phys. Rev. Lett. 31 (1973) 661--664.
\newblock \href {https://doi.org/10.1103/PhysRevLett.31.661} {\path{doi:10.1103/PhysRevLett.31.661}}.

\bibitem{Pati:1973uk}
J.~C. Pati, A.~Salam, {Unified Lepton-Hadron Symmetry and a Gauge Theory of the Basic Interactions}, Phys. Rev. D 8 (1973) 1240--1251.
\newblock \href {https://doi.org/10.1103/PhysRevD.8.1240} {\path{doi:10.1103/PhysRevD.8.1240}}.

\bibitem{Georgi:1974sy}
H.~Georgi, S.~L. Glashow, {Unity of All Elementary Particle Forces}, Phys. Rev. Lett. 32 (1974) 438--441.
\newblock \href {https://doi.org/10.1103/PhysRevLett.32.438} {\path{doi:10.1103/PhysRevLett.32.438}}.

\bibitem{Georgi:1974yf}
H.~Georgi, H.~R. Quinn, S.~Weinberg, {Hierarchy of Interactions in Unified Gauge Theories}, Phys. Rev. Lett. 33 (1974) 451--454.
\newblock \href {https://doi.org/10.1103/PhysRevLett.33.451} {\path{doi:10.1103/PhysRevLett.33.451}}.

\bibitem{Gaisser:1983vc}
T.~K. Gaisser, T.~Stanev, S.~A. Bludman, H.-s. Lee, {The Flux of Atmospheric Neutrinos}, Phys. Rev. Lett. 51 (1983) 223--226.
\newblock \href {https://doi.org/10.1103/PhysRevLett.51.223} {\path{doi:10.1103/PhysRevLett.51.223}}.

\bibitem{Barr:1988rb}
S.~M. Barr, T.~K. Gaisser, P.~Lipari, S.~Tilav, {Ratio of $\nu_e / \nu_\mu$ in Atmospheric Neutrinos}, Phys. Lett. B 214 (1988) 147--150.
\newblock \href {https://doi.org/10.1016/0370-2693(88)90468-6} {\path{doi:10.1016/0370-2693(88)90468-6}}.

\bibitem{Barr:1989ru}
G.~Barr, T.~K. Gaisser, T.~Stanev, {Flux of Atmospheric Neutrinos}, Phys. Rev. D 39 (1989) 3532--3534.
\newblock \href {https://doi.org/10.1103/PhysRevD.39.3532} {\path{doi:10.1103/PhysRevD.39.3532}}.

\bibitem{Agrawal:1995gk}
V.~Agrawal, T.~K. Gaisser, P.~Lipari, T.~Stanev, {Atmospheric neutrino flux above 1-GeV}, Phys. Rev. D 53 (1996) 1314--1323.
\newblock \href {http://arxiv.org/abs/hep-ph/9509423} {\path{arXiv:hep-ph/9509423}}, \href {https://doi.org/10.1103/PhysRevD.53.1314} {\path{doi:10.1103/PhysRevD.53.1314}}.

\bibitem{Lipari:1998cm}
P.~Lipari, T.~Stanev, T.~K. Gaisser, {Geomagnetic effects on atmospheric neutrinos}, Phys. Rev. D 58 (1998) 073003.
\newblock \href {http://arxiv.org/abs/astro-ph/9803093} {\path{arXiv:astro-ph/9803093}}, \href {https://doi.org/10.1103/PhysRevD.58.073003} {\path{doi:10.1103/PhysRevD.58.073003}}.

\bibitem{Gaisser:2001sd}
T.~K. Gaisser, {Semianalytic approximations for production of atmospheric muons and neutrinos}, Astropart. Phys. 16 (2002) 285--294.
\newblock \href {http://arxiv.org/abs/astro-ph/0104327} {\path{arXiv:astro-ph/0104327}}, \href {https://doi.org/10.1016/S0927-6505(01)00143-8} {\path{doi:10.1016/S0927-6505(01)00143-8}}.

\bibitem{Gaisser:2002zv}
T.~K. Gaisser, T.~Stanev, {Charge ratio of muons from atmospheric neutrinos}, Phys. Lett. B 561 (2003) 125--129.
\newblock \href {http://arxiv.org/abs/astro-ph/0210512} {\path{arXiv:astro-ph/0210512}}, \href {https://doi.org/10.1016/S0370-2693(03)00389-7} {\path{doi:10.1016/S0370-2693(03)00389-7}}.

\bibitem{Barr:2004br}
G.~Barr, T.~Gaisser, P.~Lipari, S.~Robbins, T.~Stanev, {A Three - dimensional calculation of atmospheric neutrinos}, Phys.Rev. D70 (2004) 023006.
\newblock \href {http://arxiv.org/abs/astro-ph/0403630} {\path{arXiv:astro-ph/0403630}}, \href {https://doi.org/10.1103/PhysRevD.70.023006} {\path{doi:10.1103/PhysRevD.70.023006}}.

\bibitem{Barr:2006it}
G.~D. Barr, S.~Robbins, T.~K. Gaisser, T.~Stanev, {Uncertainties in Atmospheric Neutrino Fluxes}, Phys. Rev. D74 (2006) 094009.
\newblock \href {http://arxiv.org/abs/astro-ph/0611266} {\path{arXiv:astro-ph/0611266}}, \href {https://doi.org/10.1103/PhysRevD.74.094009} {\path{doi:10.1103/PhysRevD.74.094009}}.

\bibitem{Gaisser:1986bv}
T.~K. Gaisser, J.~S. O'Connell, {Interactions of Atmospheric Neutrinos in Nuclei at Low-energy}, Phys. Rev. D 34 (1986) 822--825.
\newblock \href {https://doi.org/10.1103/PhysRevD.34.822} {\path{doi:10.1103/PhysRevD.34.822}}.

\bibitem{Battistoni:1999at}
G.~Battistoni, A.~Ferrari, P.~Lipari, T.~Montaruli, P.~R. Sala, T.~Rancati, {A Three-dimensional calculation of atmospheric neutrino flux}, Astropart. Phys. 12 (2000) 315--333.
\newblock \href {http://arxiv.org/abs/hep-ph/9907408} {\path{arXiv:hep-ph/9907408}}, \href {https://doi.org/10.1016/S0927-6505(99)00110-3} {\path{doi:10.1016/S0927-6505(99)00110-3}}.

\bibitem{Lipari:2000wu}
P.~Lipari, {The Geometry of atmospheric neutrino production}, Astropart. Phys. 14 (2000) 153--170.
\newblock \href {http://arxiv.org/abs/hep-ph/0002282} {\path{arXiv:hep-ph/0002282}}, \href {https://doi.org/10.1016/S0927-6505(00)00129-8} {\path{doi:10.1016/S0927-6505(00)00129-8}}.

\bibitem{Haines:1986yf}
T.~J. Haines, et~al., {Calculation of Atmospheric Neutrino Induced Backgrounds in a Nucleon Decay Search}, Phys. Rev. Lett. 57 (1986) 1986--1989.
\newblock \href {https://doi.org/10.1103/PhysRevLett.57.1986} {\path{doi:10.1103/PhysRevLett.57.1986}}.

\bibitem{Nakahata:1986zp}
M.~Nakahata, et~al., {Atmospheric Neutrino Background and Pion Nuclear Effect for Kamioka Nucleon Decay Experiment}, J. Phys. Soc. Jap. 55 (1986) 3786.
\newblock \href {https://doi.org/10.1143/JPSJ.55.3786} {\path{doi:10.1143/JPSJ.55.3786}}.

\bibitem{Hirata:1992ku}
K.~S. Hirata, et~al., {Observation of a small atmospheric muon-neutrino / electron-neutrino ratio in Kamiokande}, Phys. Lett. B 280 (1992) 146--152.
\newblock \href {https://doi.org/10.1016/0370-2693(92)90788-6} {\path{doi:10.1016/0370-2693(92)90788-6}}.

\bibitem{Fukuda:1994mc}
Y.~Fukuda, et~al., {Atmospheric muon-neutrino / electron-neutrino ratio in the multiGeV energy range}, Phys. Lett. B 335 (1994) 237--245.
\newblock \href {https://doi.org/10.1016/0370-2693(94)91420-6} {\path{doi:10.1016/0370-2693(94)91420-6}}.

\bibitem{Fukuda:1998mi}
Y.~Fukuda, et~al., {Evidence for oscillation of atmospheric neutrinos}, Phys. Rev. Lett. 81 (1998) 1562--1567.
\newblock \href {http://arxiv.org/abs/hep-ex/9807003} {\path{arXiv:hep-ex/9807003}}, \href {https://doi.org/10.1103/PhysRevLett.81.1562} {\path{doi:10.1103/PhysRevLett.81.1562}}.

\bibitem{Kajita:2000mr}
T.~Kajita, Y.~Totsuka, {Observation of Atmospheric Neutrinos}, Rev. Mod. Phys. 73 (2001) 85--118.
\newblock \href {https://doi.org/10.1103/RevModPhys.73.85} {\path{doi:10.1103/RevModPhys.73.85}}.

\bibitem{Ambrosio:1998wu}
M.~Ambrosio, et~al., {Measurement of the atmospheric neutrino induced upgoing muon flux using MACRO}, Phys. Lett. B 434 (1998) 451--457.
\newblock \href {http://arxiv.org/abs/hep-ex/9807005} {\path{arXiv:hep-ex/9807005}}, \href {https://doi.org/10.1016/S0370-2693(98)00885-5} {\path{doi:10.1016/S0370-2693(98)00885-5}}.

\bibitem{Honda:1995hz}
M.~Honda, T.~Kajita, K.~Kasahara, S.~Midorikawa, {Calculation of the flux of atmospheric neutrinos}, Phys. Rev. D 52 (1995) 4985--5005.
\newblock \href {http://arxiv.org/abs/hep-ph/9503439} {\path{arXiv:hep-ph/9503439}}, \href {https://doi.org/10.1103/PhysRevD.52.4985} {\path{doi:10.1103/PhysRevD.52.4985}}.

\bibitem{Honda:2006qj}
M.~Honda, T.~Kajita, K.~Kasahara, S.~Midorikawa, T.~Sanuki, {Calculation of atmospheric neutrino flux using the interaction model calibrated with atmospheric muon data}, Phys.Rev. D75 (2007) 043006.
\newblock \href {http://arxiv.org/abs/astro-ph/0611418} {\path{arXiv:astro-ph/0611418}}, \href {https://doi.org/10.1103/PhysRevD.75.043006} {\path{doi:10.1103/PhysRevD.75.043006}}.

\bibitem{Gaisser:2002jj}
T.~K. Gaisser, M.~Honda, {Flux of atmospheric neutrinos}, Ann. Rev. Nucl. Part. Sci. 52 (2002) 153--199.
\newblock \href {http://arxiv.org/abs/hep-ph/0203272} {\path{arXiv:hep-ph/0203272}}, \href {https://doi.org/10.1146/annurev.nucl.52.050102.090645} {\path{doi:10.1146/annurev.nucl.52.050102.090645}}.

\bibitem{Gaisser:2013bla}
T.~K. Gaisser, T.~Stanev, S.~Tilav, {Cosmic Ray Energy Spectrum from Measurements of Air Showers}, Front. Phys.(Beijing) 8 (2013) 748--758.
\newblock \href {http://arxiv.org/abs/1303.3565} {\path{arXiv:1303.3565}}, \href {https://doi.org/10.1007/s11467-013-0319-7} {\path{doi:10.1007/s11467-013-0319-7}}.

\bibitem{Gaisser:1977yra}
T.~K. Gaisser, A.~Halprin, {Detection of very high energy neutrinos}, in: {15th International Cosmic Ray Conference}, 1977.

\bibitem{Gaisser:1978mj}
T.~K. Gaisser, A.~Halprin, {Higgs Meson Production in Muon Number Violating Processes}, Phys. Rev. D 17 (1978) 159--162.
\newblock \href {https://doi.org/10.1103/PhysRevD.17.159} {\path{doi:10.1103/PhysRevD.17.159}}.

\bibitem{Elbert:1980sq}
J.~W. Elbert, T.~K. Gaisser, T.~Stanev, {Possible Studies with DUMAND and a Surface Air Shower Detector.}, in: {1980 DUMAND Symposium}, 1980.

\bibitem{Gaisser:1979hp}
{Proceedings, Topical Conference on Cosmic Rays and Particle Physics}: {Newark, Delaware, October 16-21, 1978}, Vol.~49 of Particles and Fields Subseries.

\bibitem{Samorski:1983zm}
M.~Samorski, W.~Stamm, {Detection of 2 x 10**15-eV to 2 x 10**16-eV gamma Rays from Cygnus X-3}, Astrophys. J. Lett. 268 (1983) L17--L21.
\newblock \href {https://doi.org/10.1086/184021} {\path{doi:10.1086/184021}}.

\bibitem{Abdo:2009kfa}
A.~A. Abdo, et~al., {Modulated High-Energy Gamma-Ray Emission from the Microquasar Cygnus X-3}, Science 326~(5959) (2009) 1512--1516.
\newblock \href {https://doi.org/10.1126/science.1182174} {\path{doi:10.1126/science.1182174}}.

\bibitem{Gaisser:1983cj}
T.~K. Gaisser, T.~Stanev, {Calculation of Neutrino Flux From Cygnus X-3}, Phys. Rev. Lett. 54 (1985) 2265.
\newblock \href {https://doi.org/10.1103/PhysRevLett.54.2265} {\path{doi:10.1103/PhysRevLett.54.2265}}.

\bibitem{BaldoCeolin:1990hg}
M.~Baldo~Ceolin (Ed.), {Neutrino telescopes. Proceedings, 2nd International Workshop, Venice, Italy, February 13-15, 1990}, 1990.

\bibitem{Gaisser1995a}
T.~K. Gaisser, F.~Halzen, T.~Stanev, {Particle astrophysics with high-energy neutrinos}, Phys.Rept. 258 (1995) 173--236.
\newblock \href {http://arxiv.org/abs/hep-ph/9410384} {\path{arXiv:hep-ph/9410384}}, \href {https://doi.org/10.1016/0370-1573(95)00003-Y} {\path{doi:10.1016/0370-1573(95)00003-Y}}.

\bibitem{Gaisser:1987fn}
T.~K. Gaisser, T.~Stanev, F.~Halzen, {LIMITS ON THE ULTRAHIGH-ENERGY RADIATION BY YOUNG SUPERNOVAE} (10 1987).

\bibitem{Gaisser:1987wm}
T.~K. Gaisser, T.~Stanev, {Energetic (\ensuremath{>} {GeV}) Neutrinos as a Probe of Acceleration in the New Supernova}, Phys. Rev. Lett. 58 (1987) 1695, [Erratum: Phys.Rev.Lett. 59, 844 (1987)].
\newblock \href {https://doi.org/10.1103/PhysRevLett.58.1695} {\path{doi:10.1103/PhysRevLett.58.1695}}.

\bibitem{Gaisser:1986ha}
T.~K. Gaisser, G.~Steigman, S.~Tilav, {Limits on Cold Dark Matter Candidates from Deep Underground Detectors}, Phys. Rev. D 34 (1986) 2206.
\newblock \href {https://doi.org/10.1103/PhysRevD.34.2206} {\path{doi:10.1103/PhysRevD.34.2206}}.

\bibitem{Gaisser:1991tz}
T.~K. Gaisser, T.~Stanev, F.~Halzen, W.~F. Long, E.~Zas, {Gamma-ray astronomy above 50-TeV with muon poor showers}, Phys. Rev. D 43 (1991) 314--318.
\newblock \href {https://doi.org/10.1103/PhysRevD.43.314} {\path{doi:10.1103/PhysRevD.43.314}}.

\bibitem{Seckel:1990rk}
D.~Seckel, T.~Stanev, T.~K. Gaisser, {Albedo gamma-ray from cosmic ray interactions on the solar surface}, AIP Conf. Proc. 220 (1991) 227--231.
\newblock \href {https://doi.org/10.1063/1.40299} {\path{doi:10.1063/1.40299}}.

\bibitem{Arguelles:2019jfx}
C.~A. Arg\"uelles, A.~Kheirandish, J.~Lazar, Q.~Liu, {Search for Dark Matter Annihilation to Neutrinos from the Sun}, PoS ICRC2019 (2021) 527.
\newblock \href {http://arxiv.org/abs/1909.03930} {\path{arXiv:1909.03930}}, \href {https://doi.org/10.22323/1.358.0527} {\path{doi:10.22323/1.358.0527}}.

\bibitem{Ajello:2021agj}
M.~Ajello, et~al., {First Fermi-LAT Solar Flare Catalog}, Astrophys. J. Suppl. 252~(2) (2021) 13, [Erratum: Astrophys.J.Suppl. 256, 24 (2021)].
\newblock \href {http://arxiv.org/abs/2101.10010} {\path{arXiv:2101.10010}}, \href {https://doi.org/10.3847/1538-4365/abd32e} {\path{doi:10.3847/1538-4365/abd32e}}.

\bibitem{Gaisser:1985cm}
T.~K. Gaisser, T.~Stanev, {Response of Deep Detectors to Extraterrestrial Neutrinos}, Phys. Rev. D 31 (1985) 2770.
\newblock \href {https://doi.org/10.1103/PhysRevD.31.2770} {\path{doi:10.1103/PhysRevD.31.2770}}.

\bibitem{Gaisser:1998hb}
T.~K. Gaisser, {Fluxes of atmospheric neutrinos and related cosmic rays}, Nucl. Phys. B Proc. Suppl. 77 (1999) 133--139.
\newblock \href {http://arxiv.org/abs/hep-ph/9811315} {\path{arXiv:hep-ph/9811315}}, \href {https://doi.org/10.1016/S0920-5632(99)00408-9} {\path{doi:10.1016/S0920-5632(99)00408-9}}.

\bibitem{Barr:2007fza}
G.~Barr, T.~K. Gaisser, T.~Stanev, {Uncertainty Estimates for Atmosheric Neutrino Fluxes}, in: {30th International Cosmic Ray Conference}, Vol.~5, 2007, pp. 1495--1498.

\bibitem{Aartsen:2013jdh}
M.~G. Aartsen, et~al., {Evidence for High-Energy Extraterrestrial Neutrinos at the IceCube Detector}, Science 342 (2013) 1242856.
\newblock \href {http://arxiv.org/abs/1311.5238} {\path{arXiv:1311.5238}}, \href {https://doi.org/10.1126/science.1242856} {\path{doi:10.1126/science.1242856}}.

\bibitem{Aartsen:2014gkd}
M.~G. Aartsen, et~al., {Observation of High-Energy Astrophysical Neutrinos in Three Years of IceCube Data}, Phys. Rev. Lett. 113 (2014) 101101.
\newblock \href {http://arxiv.org/abs/1405.5303} {\path{arXiv:1405.5303}}, \href {https://doi.org/10.1103/PhysRevLett.113.101101} {\path{doi:10.1103/PhysRevLett.113.101101}}.

\bibitem{IceCube:2022der}
R.~Abbasi, et~al., {Evidence for neutrino emission from the nearby active galaxy NGC 1068}, Science 378~(6619) (2022) 538--543.
\newblock \href {http://arxiv.org/abs/2211.09972} {\path{arXiv:2211.09972}}, \href {https://doi.org/10.1126/science.abg3395} {\path{doi:10.1126/science.abg3395}}.

\bibitem{IceCube:2018dnn}
M.~G. Aartsen, et~al., {Multimessenger observations of a flaring blazar coincident with high-energy neutrino IceCube-170922A}, Science 361~(6398) (2018) eaat1378.
\newblock \href {http://arxiv.org/abs/1807.08816} {\path{arXiv:1807.08816}}, \href {https://doi.org/10.1126/science.aat1378} {\path{doi:10.1126/science.aat1378}}.

\bibitem{IceCube:2018cha}
M.~G. Aartsen, et~al., {Neutrino emission from the direction of the blazar TXS 0506+056 prior to the IceCube-170922A alert}, Science 361~(6398) (2018) 147--151.
\newblock \href {http://arxiv.org/abs/1807.08794} {\path{arXiv:1807.08794}}, \href {https://doi.org/10.1126/science.aat2890} {\path{doi:10.1126/science.aat2890}}.

\bibitem{yu2023recent}
S.~Yu, J.~Micallef, Recent neutrino oscillation result with the icecube experiment (2023).
\newblock \href {http://arxiv.org/abs/2307.15855} {\path{arXiv:2307.15855}}.

\end{thebibliography}
\end{document}